\documentclass[a4paper,11pt]{article}
\pdfoutput=1 

\usepackage{jcappub}

\usepackage[T1]{fontenc} 
\usepackage{comment}

\newcommand{\dis}[1]{\begin{equation}\begin{split}#1\end{split}\end{equation}}

\newcommand{\bfrac}[2]{\left(\frac{#1}{#2} \right)  }
\usepackage{xcolor,cancel}

\title{\boldmath Gravitational Wave Sourced by Decay of Massive Particle from Primordial Black Hole evaporation}

\author[a,b]{Ki-Young Choi,}
\author[a]{Erdenebulgan Lkhagvadorj,}
\author[a]{and Satyabrata Mahapatra}

\affiliation[a]{Department of Physics and Institute of Basic Science, Sungkyunkwan University, 2066 Seobu-ro, Suwon-si, Gyeonggi-do, 16419, Korea}
\affiliation[b]{Korea Institute for Advanced Study, Seoul 02455, Korea}

\emailAdd{kiyoungchoi@skku.edu}
\emailAdd{bulgaa@skku.edu}
\emailAdd{satyabrata@skku.edu}

\abstract{ In this article, we investigate the stochastic gravitational waves (GWs) spectrum, resulting from the emission of gravitons through bremsstrahlung, in the decay of particles produced by Hawking radiation. Although particle decays inevitably entail the emission of graviton due to bremsstrahlung, the associated decay width is notably suppressed due to the Planck scale suppression in the coupling of matter fields to gravitons. Consequently, the relic abundance of such GWs constituted of these gravitons undergoes a corresponding reduction. However, we demonstrate that super-heavy particles, reaching masses as high as Planck scale, can emerge naturally in the Hawking radiation of evaporating primordial black holes (PBHs) and can compensate for this suppression. In addition, we also discuss the stochastic gravitational waves constituted out of the gravitons directly radiated from such evaporating PBHs. When the super-heavy particle decays promptly after its production, then the corresponding GW spectrum remains subdominant to the one arising from direct PBH evaporation. However, if this particle is long-lived and decays after PBH evaporation, then the resulting GWs produced in these two processes have two distinct spectra with their peaks at extremely high frequencies, providing avenues for proposed ultra-high frequency gravitational wave detectors. We also show that such gravitational waves contribute significantly to substantial dark radiation, which can be probed with the enhanced sensitivity of future experiments.  }

\hypersetup{
colorlinks = true,
linkcolor = blue,
citecolor = magenta
}

\begin{document}
\maketitle
\flushbottom

\section{Introduction}\label{intro}

The discovery of Gravitational Waves (GWs) by the LIGO and VIRGO collaboration~\cite{LIGOScientific:2016aoc,LIGOScientific:2017vwq} from different astrophysical sources as well as the detection of stochastic gravitational wave background (SGWB), which might be potentially of cosmological origin, by different Pulsar Timing Array (PTA) collaborations~\cite{NANOGrav:2023gor,EPTA:2023fyk,Reardon:2023gzh,Xu:2023wog}, have ignited a great intrigue among the particle physics and cosmology community. The discovery of GWs  hold profound significance as GWs can traverse the Universe practically unimpeded unlike the electromagnetic radiation. Thus, GWs can carry undistorted information about their source and hence they are the purest probes of physical phenomena even at exceptionally high energy scales. Out of many possible phenomena that can source such GWs, Hawking evaporation of primordial black holes (PBHs)~\cite{Hawking:1974rv,Hawking:1975vcx} that may have formed in the early Universe is of great importance, as the ongoing elusiveness of any clear, non-gravitational signals from any dark matter(DM) candidates, has reignited interest in exploring PBHs, which can be potential DM candidates themselves or play a crucial role in DM production (see e.g.~\cite{Gondolo:2020uqv,Bernal:2020bjf} and refrences therein). PBHs have also been explored to address Baryogenesis~\cite{Carr:1976zz,Baumann:2007yr,Hook:2014mla,Hamada:2016jnq,Choi:2023kxo} as well as in Cogenesis scenarios trying to address the Baryon-DM coincidence problem (see e.g.~\cite{Morrison:2018xla,Smyth:2021lkn, Barman:2022pdo}).

Black holes are theorized to emit quasi-thermal radiation, known as Hawking radiation, which leads to their gradual evaporation over time~\cite{Hawking:1974rv,Hawking:1975vcx}. This process allows black holes to dissipate energy at predictable rates into all physical degrees of freedom that have a mass at or below the black hole's temperature~\cite{Page:1976ki}. 
The implications of this evaporation extend far beyond the fate of the black holes themselves. As they radiate away their mass, PBHs could potentially unleash a cascade of particles, of both visible sector and the dark sector. Moreover, the gravitons emitted in the Hawking evaporation can not equilibrate and thus survive until the present day, generating a stochastic background of gravitational waves of high frequency ~\cite{Anantua_2009, Dolgov:2011cq, Dong_2016, Ireland_2023}. This offers a unique signature that, if detected, could provide unprecedented insights into the early Universe.
The prospect of detecting such gravitational waves is particularly fascinating as it can potentially be sourced from light PBHs with masses below $\sim 10^9$ g, which would have evaporated before the era of big bang nucleosynthesis (BBN), leaving no other observable traces~\cite{Carr:2021bzv, Escriva:2022duf}.

In addition, there is a possibility that extremely heavy particles with sub-Planckian mass could be produced during Hawking evaporation, particularly towards the end of PBH evaporation when the PBH temperature is extremely high. These super-heavy particles could have significant implications and unique detection prospects, which are the central focus of our discussion in this article. Motivations for such super-heavy particles in various particle physics models include their potential role as dark matter candidates, involvement in baryogenesis processes, association with inflationary models, addressing the hierarchy problem etc. These particles, if produced in Hawking evaporation, can decay into SM particles or particles of the dark sector, potentially generating gravitons through bremsstrahlung. 

Gravitational waves originating from such quantum gravitational processes have been studied in some recent works involving scattering of particles~\cite{Ghiglieri:2020mhm,Ringwald:2020ist}, particle annihilation to gravitons~\cite{Ghiglieri:2022rfp,Choi:2024ilx} and bremsstrahlung in particle decays~~\cite{Nakayama:2018ptw,Huang:2019lgd,Barman:2023ymn, Bernal:2023wus, Barman:2023rpg, Kanemura:2023pnv, Hu:2024awd}. Given that particle decays are invariably accompanied by graviton emission, gravitational waves stemming from graviton bremsstrahlung hold particular significance among these quantum gravitational phenomena.
However production of gravitons through such decays is typically suppressed due to the coupling of matter fields to gravitons by the Planck mass (the effective interaction can be written as $\mathcal{L}_{\rm eff} \supset \lambda h_{\mu\nu}T^{\mu\nu}$ with $\lambda \sim M^{-1}_p$ where $h_{\mu\nu}$ is the graviton field and $T^{\mu\nu}$ is the energy momentum tensor of other matter fields). But in the case of extremely heavy sub-Planckian mass mother particles, this suppression can be compensated. This compensation could lead to a sufficient energy transfer to gravitons, resulting in observable ultra high frequency gravitational wave signatures.

Here, we study the GW stemming from the graviton production in Hawking evaporation as well as GW  production through bremsstrahlung process in the decay of a long-lived super-heavy scalar produced in PBH evaporation.
We illustrate that GW production through bremsstrahlung can have a remarkable enhancement over a wide frequency range, without any suppression, being abundantly sourced from the super-heavy particle.  
We find that resulting GWs produced in these two processes have two distinct spectra, thus offering a potential avenue for probing Planck-scale physics, an endeavor that is otherwise exceedingly challenging.  
These spectra of GWs from both sources peak at extremely high frequencies. We demonstrate that the peak frequency for the spectra of the GW from PBH evaporation is decided by the PBH initial mass, whereas that for the GWs from graviton bremsstrahlung in particle decay is decided by the mass of the super-heavy particle and the corresponding coupling involved.  We show that the resonant cavity detectors~\cite{Herman:2022fau, Berlin:2021txa, Berlin:2023grv} have the potential to detect GW signal. Also such GWs can contribute to testable dark radiation that can be constrained by the CMB observations~\cite{Planck:2018vyg} and can be detected in future CMB experiments with enhanced sensitivities~\cite{CMB-S4:2016ple,laureijs2011euclid,Ben-Dayan:2019gll}.

The manuscript is built up as follows: we begin by discussing the formation and evaporation characteristics of PBHs in Section \ref{section2}, followed by the discussion of gravitational wave generation resulting from Hawking radiation of gravitons in Section \ref{section3}. Subsequently, in Section \ref{section4}, we calculate the abundance of long-lived super-heavy particles generated during PBH evaporation and explore the production of ultra-high-frequency gravitational waves from the graviton bremsstrahlung in their decay and finally conclude in Section \ref{section5}. 

\section{Primordial Black Hole: Generation and Evaporation} \label{section2}
Primordial black holes might have formed in the early universe through various mechanisms, distinct from those that create stellar black holes. These mechanisms often involve the gravitational collapse of matter overdensities, which could have been seeded by inflation or resulted from topological defects. 
The concept of PBH formation was first proposed by Zeldovich and Novikov~\cite{Zeldovich:1967lct} in the late 1960s and was further developed by Hawking and Carr in the early 1970s~\cite{Carr:1974nx}.  PBHs are hypothesized to have formed during the radiation-dominated epoch immediately following the inflationary era \cite{Carr_2010, Carr_2021, Carr:2020xqk}.
In the radiation-dominated Universe, the initial mass of black hole denoted as $M_{\rm in}$, at the moment of its formation can be linked to the background energy density $\rho$ and the Hubble parameter $H$ at the corresponding temperature $T_{\rm in}$ as
~\cite{Fujita2014, Masina:2020xhk, Gondolo:2020uqv}
\begin{equation}\label{eq:MassPBH}
    M_{\rm in} \equiv M_{\rm BH}(T_{\text{in}}) = \gamma \frac{4 \pi}{3} \frac{\rho(T_{\rm in})}{H^3(T_{\rm in})},
\end{equation}
   where $\gamma\simeq 0.2$ is a numerical correction factor, $\rho(T_{\rm in}) = 3 M_p^2 H^2(T_{\rm in})$ is the total energy density. Here, $M_p = 1/\sqrt{8\pi G}\simeq 2.4 \times 10^{18} \ \rm{GeV}$ is the reduced Planck mass. As a result, the temperature of the plasma, denoted as $T_{\rm in}$ can be expressed in relation to $M_{\rm in}$ as
\begin{equation}\label{eq:initialTem}
    T_{\rm in} = \left(\frac{1440 \ \gamma^2}{g_*(T_{\rm in})}\right)^{1/4} M_p \sqrt{\frac{M_p}{M_{\rm in}}} \simeq 4.36 \times 10^{15} \ \text{GeV} \left(\frac{1 \ \rm{g}}{M_{\rm in}}\right)^{1/2},
\end{equation}
where $g_*(T_{\rm in})$ denotes the effective degrees of freedom of the energy density at the formation temperature and ${1\ {\rm g}\simeq 2.34 \times 10^5 M_p}$. Once formed, PBH can evaporate  by emitting Hawking radiation, transitioning into particles, if their masses are smaller than the instantaneous Hawking temperature of PBH, $T_{\text{BH}}$. For Schwarzschild black hole, $T_{\text{BH}}$ is  given by~\cite{Hawking:1974rv, Hawking:1975vcx} 
\begin{equation}\label{eq:Hawkingtemperature}
    T_\text{BH} = \frac{M_p^2}{M_\text{BH}}\simeq 
    10^{13} \ \rm GeV \left(\frac{1 \ \rm g}{M_{\rm BH}}\right).
\end{equation}
In case of Schwarzschild black holes, the 
energy spectrum of the i-th species evaporated from the PBH can be written as~\cite{UKWATTA201690, Lunardini:2019zob, Perez-Gonzalez:2020vnz}:
\begin{equation}\label{eq:energy_spectrum}
    \frac{d^2u_i(E,t)}{dE dt} = \frac{g_i}{2 \pi^2} \frac{\sigma_{s_i}(M_\text{BH}, \mu_i,E_i)}{e^{E_i/T_\text{BH}} -(-1)^{2s_i}} E_i^3, 
\end{equation}
where $g_i$ denotes the number of degrees of freedom of $i$-th species, $E_i = \sqrt{\mu_i^2 + p^2}$ represents the total energy of particles with mass $\mu_i$ and spin $s_i$, and $\sigma_{s_i}$ signifies the absorption cross section of a state with energy $E$ by a single PBH. Thus, the rate of mass loss for the PBH can be obtained by 
summing the energy spectrum equation, Eq.~(\ref{eq:energy_spectrum}), over all possible particle species, and integrating over the phase space, which can be written as follows~\cite{Cheek2022}:
\begin{equation}
\begin{aligned}[b]
   \frac{dM_\text{BH}}{dt} &
   = - \sum_i \int_0^\infty \frac{d^2u_i(E,t)}{dE dt} dE  = - \varepsilon(M_\text{BH}) \frac{  M^4_p}{M_\text{BH}^2},
\end{aligned}
\end{equation}
where $\varepsilon(M_\text{BH}) \equiv \sum_i g_i \varepsilon_i (z_i)$ is defined as an evaporation function with the function $\varepsilon_i (z_i)$ given by~\cite{Cheek2022} 
\begin{equation}\label{eq:evap Func}
    \varepsilon_i (z_i) =\frac{27}{128\pi^3} \int_{z_i}^\infty \frac{\psi_{s_i}(x) (x^2 - z_i^2)}{e^x - (-1)^{2s_i}} x dx.
\end{equation}
Here  $x=E_i/T_\text{BH}$ and $z_i = \mu_i/T_\text{BH}$ with $T_\text{BH}$ given in Eq.~(\ref{eq:Hawkingtemperature}) and we define $\psi_{s_i}(E) = \sigma_{s_i}(E)/\left(\frac{27}{64\pi} \frac{M^2_{\rm BH}}{M^4_p}\right)$.

In the geometric optics limit, where $\psi_{s_i}(E)=1$, and neglecting the mass of the particles, the mass loss rate~\cite{Baldes_2020, Cheek2022} can be approximated as
\begin{equation}\label{eq:mass loss}
    \frac{d M_\text{BH}}{dt} \simeq - \frac{27\pi}{4} \frac{  \ g_*(T_\text{BH})}{480} \frac{M_p^4}{M_\text{BH}^2},
\end{equation}
where $g_*(T_{\rm BH})$ is the total number of relativistic degrees of freedom emitted by PBH.
By solving Eq.~(\ref{eq:mass loss}), the time evolution of PBH mass is given by
\begin{equation}\label{eq:PBHmass}
    M_\text{BH}(t) = M_\text{in} \left(1- \frac{t-t_{i}}{\tau}\right)^{1/3},
\end{equation}
with the lifetime $\tau$ as
\begin{eqnarray}\label{eq:lifetime}
    \tau = \frac{4}{27} \frac{160 \ M_\text{in}^3}{\pi \ g_*(T_\text{BH}) M_p^4} \simeq 2.66\times 10^{-28} \ \rm{s} \bfrac{100}{g_*(T_\text{BH})}\left(\frac{M_{\rm in}}{1 \ \rm{g}}\right)^{3}.
\end{eqnarray}
The majority of the energy loss from PBH occurs during the final stages of the evaporation process, when its mass has diminished significantly.
The evaporation temperature, denoted as $T_\text{ev}$, of the background plasma immediately following the evaporation of the PBH, can be characterized at the cosmic time $\tau$ by setting $\rho_\text{rad}(T_\text{ev})= 3M_p^2 H^2(\tau)$. When the PBH dominates the energy density of the Universe prior to its evaporation, the Universe enters a matter-dominated phase. In such a scenario, $H^2=\frac{4}{9 \tau^2}$, resulting in the evaporation temperature given by~\cite{Masina:2020xhk}:

\dis{ \label{eq:evapTemperature}
  T_\text{ev} |_\text{MD} &\equiv \left( \frac{40 M_p^2}{\pi^2 g_*(T_\text{ev})\tau^2}\right)^{1/4}\simeq {0.5}\bfrac{g_*^2(T_{\rm BH})}{g_*( T_{\rm ev})}^{1/4}\bfrac{M_p^5}{M_{\rm in}^3}^{1/2}
  \simeq {3.55 \times 10^{10}} \ \text{GeV} \left(\frac{1 \ \text{g}}{M_\text{in}}\right)^{3/2},
}
where $g_*(T_\text{BH})\simeq g_*(T_\text{ev})\simeq 100$ is used.
In scenarios where PBH are subdominant before evaporation, $H^2(T_\text{ev}) = \frac{1}{4 \tau^2}$ can be utilized leading to an evaporation temperature which is slightly smaller than that in the matter-dominated phase~\cite{Masina:2020xhk}
\begin{equation}\label{eq:ev temperaturePBH}
    T_\text{ev} |_\text{RD} \simeq \frac{\sqrt{3}}{2} T_\text{ev} |_\text{MD}.
\end{equation}

A lower limit on the mass of PBH is imposed by the maximum Hubble size of the Universe following inflation, as the tensor-to-scalar ratio is constrained by observations of cosmic microwave background anisotropy. This constraint is expressed as $H < 2.5 \times 10^{-5} M_p$~\cite{Planck:2018jri}.
Moreover, when PBH dominates the Universe before its evaporation, it is crucial that the evaporation temperature is sufficiently high to avoid disrupting the predictions of big bang nucleosynthesis (BBN), necessitating $T_\text{ev} > T_\text{BBN}\simeq 4 \ \text{MeV}$. This requirement sets an upper bound on the PBH mass. By considering both these constraints, we determine the available range of PBH mass as:
\begin{equation}\label{eq:PBH mass upper limit}
0.4 \ \text{g} \lesssim M_\text{in} \lesssim 9.7 \times 10^8 \ \text{g}.
\end{equation}

\section{Gravitational Wave production from PBH evaporation}\label{section3}
As explained in the previous section, PBH can produce a wide array of particles with masses $m\lesssim T_{\rm BH}$, including gravitons, through Hawking evaporation.
For the sake of simplicity, considering a Schwarzschild black hole, the greybody factor, which is a measure of the efficiency of particle emission, can be approximated as $\sigma_{s_i}=\left(\frac{27}{64\pi} \frac{M^2_{\rm BH}}{M^4_p}\right)$ in the geometric optics limit. 
Consequently, taking into account the two graviton polarizations $g_i = 2$ and the spin of graviton $s=2$, Eq.~(\ref{eq:energy_spectrum}) simplifies to the following expression for graviton production from the direct evaporation of PBH;
\begin{equation}
    \frac{d^2 u_{\rm GW}}{dtdE} \simeq \frac{27}{64 \pi^3} \frac{M_{\rm BH}^2}{M_p^4} \frac{E^3}{e^{E/T_{\rm BH}}-1}.
\end{equation}
Therefore, calculating the rate of graviton emission for a collective population of evaporating PBHs, we can write
\begin{equation}
    \frac{d\rho_{\rm GW}}{dt dE} \simeq n_{\rm BH} (t) \frac{d^2u_{\rm GW}}{dt dE}. 
\end{equation}
where $n_{\rm BH}(t)$ is the PBH number density. 
Here, it should be worth keeping in mind that, the presence of PBH alongside the standard radiation bath contributes to the overall density and alters the Hubble expansion rate $H$ of the early universe expressed as:
\begin{equation}
    H^2 = \frac{1}{3 M_p^2} (\rho_r + \rho_\text{BH}).
\end{equation}
Using the comoving energy density of radiation and PBH 
defined as $\tilde{\rho}_r \equiv a^4 \times \rho_r$ and $\tilde{\rho}_{\rm BH} \equiv a^3 \times \rho_{\rm BH}$, 
where $a$ represents the scale factor, the Boltzmann equations governing the evolution of the Universe are given by~\cite{Masina:2020xhk, Giudice:2000ex, Bernal:2020bjf, JyotiDas:2021shi, Barman:2021ost}: 
\begin{equation}
\begin{aligned}[b]
    & \frac{dM_\text{BH}}{d \ln(a)} = - \frac{\varepsilon(M_\text{BH})}{H} \frac{ M_p^4}{M_\text{BH}^2}, \\
    & \frac{d \tilde{\rho}_\text{BH}}{d \ln(a)} =  \frac{\tilde{\rho}_\text{BH}}{M_\text{BH}} \frac{dM_\text{BH}}{d \ln(a)}, \\
    & \frac{d \tilde{\rho}_r}{d \ln(a)} = -\frac{\varepsilon_\text{SM}(M_\text{BH})}{\varepsilon(M_\text{BH})} \frac{a \ \tilde{\rho}_\text{BH}}{M_\text{BH}} \frac{dM_\text{BH}}{d \ln(a)}\,.
     \label{Bolteq}
\end{aligned}
\end{equation}
Here $\varepsilon_\text{SM}\equiv g_{\rm SM}\sum_i \varepsilon_i (z_i)$ represents the contribution solely from the SM to the  evaporation excluding the contribution from gravitons. 
In this paper, we assume a monochromatic mass spectrum for the PBH and thus the comoving number density of PBH remain conserved, {\it i.e.} $\tilde{n}_{\rm BH}$=constant. Hence, the PBH number density at any time can be related to the initial number density $n_{{\rm BH},i}$ as:
\begin{equation}\label{eq:scaling numden}
    n_{\rm BH} (t) = n_{\text{BH}, i} \left(\frac{a_i}{a}\right)^3.
\end{equation}
Therefore, the energy density of the PBHs can be expressed as 
\begin{equation}\label{eq:PBH energy density}
    \rho_{\rm BH}(t) = n_{\rm BH}(t) M_{\rm BH}(t) = \rho_{\rm BH}(t_{\rm in}) \left(\frac{a_i}{a}\right)^3
    \left(1- \frac{t-t_{i}}{\tau}\right)^{1/3}.
\end{equation}
{Using sudden decay approximation with $\rho_{\rm BH}(t_{\rm ev}) = \rho_r(t_{\rm ev})$, we find the relation,
\dis{
\rho_{\rm BH}(t_{\rm ev})= \rho_{\rm BH}(t_i) \bfrac{a_i}{a_{\rm ev}}^3= \rho_r(t_{\rm ev}) = \frac{\pi^2}{30}g_*(T_{\rm ev}) T_{\rm ev}^4
}
Therefore,
\dis{\label{eq:nbhiai3}
n_{\text{BH},i} a_i^3 = \bfrac{\rho_r(t_{\rm ev}) a_{\rm ev}^3}{M_{\rm in}}=\bfrac{ \frac{\pi^2}{30}g_*(T_{\rm ev}) T_{\rm ev}^4 a_{\rm ev}^3}{M_{\rm in}}.
}
}
Following their production, the energy density and frequency of GW undergo redshifting, and their subsequent values can be related to those at the time of complete PBH evaporation by the equations:
\begin{equation}\label{eq:scalingfor rho frequency}
    \rho_{\rm GW} = \rho_{\rm GW, ev} \left(\frac{a_{\rm ev}}{a}\right)^4, \quad \omega = \omega_{\rm ev} \left(\frac{a_{\rm ev}}{a}\right).
\end{equation}
Utilizing  Eqs. (\ref{eq:scaling numden}) and (\ref{eq:scalingfor rho frequency}) for the scaling as well as Eq.~(\ref{eq:PBHmass}) for the PBH mass evolution, the energy density for gravitational waves at the time of PBH evaporation can be estimated as:
\begin{equation}
    \frac{d \rho_{\rm GW, ev}}{d \ln \omega_{\rm ev}} = \frac{27}{64 \pi^3} \frac{M_{\rm in}^2}{M_p^4} n_{\rm BH}(t_i) \omega_{\rm ev}^4 \int_{t_i}^{t_{\rm ev}=t_i+\tau} dt \left(1-\frac{t-t_i}{\tau}\right)^{2/3} \frac{(a_i/a)^3}{e^{\omega_{\rm ev}a_{\rm ev}/a T_{\rm BH}}-1},
\end{equation}
The relationship between the energy density of gravitational waves today and that at the time of evaporation is expressed as:
\begin{equation}
    \frac{d\rho_{\rm GW,0}}{d \ln \omega_0} = \frac{d \rho_{\rm GW, ev}}{d \ln \omega_{\rm ev}} \left(\frac{a_{\rm ev}}{a_0}\right)^4,
\end{equation}
with
\begin{equation}\label{eq:aevtoa0}
     \frac{a_{\rm ev}}{a_{0}} \simeq \frac{T_{0}}{T_{\rm ev}} \left(\frac{g_{*,s}(T_{0})}{g_{*,s}(T_{\rm ev})}\right)^{1/3} 
    \simeq {6.2 \times 10^{-32}} \left(\frac{g_{*,s}(T_{0})}{g_{*,s}(T_{\rm ev})}\right)^{1/3} \left(\frac{M_{\rm in}}{M_p}\right)^{3/2} 
    \simeq 2.3 \times 10^{-24 } \left(\frac{M_{\rm in}}{1 \ \rm{g}}\right)^{3/2},
\end{equation}
where we take $T_0 \simeq 2.3 \times 10^{-13} \ \rm{GeV}$, $g_{*,s}(T_0)=3.91$ and $g_*(T_\text{ev}) \simeq g_{*,s}(T_\text{ev}) \simeq 100$.  
Denoting the redshifted frequency  values today as $\omega_0= \omega_{\rm ev} a_{\rm ev}$, with $a_0=1$, the energy density of gravitational waves at present can be expressed as 
\begin{eqnarray}\label{eq:rhoGW0}
    \frac{d\rho_{\rm GW,0}}{d \ln \omega_0} &=& \frac{27}{64 \pi^3} \frac{M_{\rm in}^2}{M_p^4} n_{\rm BH}(t_i) \omega_{0}^4 \int_{t_i}^{t_{\rm ev}} dt \left(1-\frac{t-t_i}{\tau}\right)^{2/3} \frac{(a_i/a)^3}{e^{\omega_{0}/a T_{\rm BH}}-1} \nonumber \\
    &\simeq& {2.8} \times 10^{-4} g_*^2(T_{\rm BH}) \frac{M_p^6}{M_{\rm in}^5 } \omega_0^4  \int_{t_i}^{t_{\rm ev}} dt \left(1-\frac{t-t_i}{\tau}\right)^{2/3} \frac{(a_{ \rm ev}/a)^3}{e^{\omega_{0}/a T_{\rm BH}}-1},
\end{eqnarray}
where in the last equality, we utilize the fact that the number density of PBH remains constant over time intervals evident from Eq.~(\ref{eq:nbhiai3}). Using Eq.~(\ref{eq:aevtoa0}) in Eq.~(\ref{eq:rhoGW0}), we obtain:
\begin{eqnarray}
     \frac{d\rho_{\rm GW,0}}{d \ln \omega_0} 
     &\simeq& {6.2 \times 10^{-77}} \ {\rm GeV} \left(\frac{M_p}{M_{\rm in}}\right)^{1/2}  \omega_0^4 \ I(\omega_0), 
\end{eqnarray}
where the frequency dependent integral is defined as
\begin{equation} \label{eq:Iw0}
    I(\omega_0) = \int_{t_i}^{t_{\rm ev}} dt \left(1-\frac{t-t_i}{\tau}\right)^{2/3} \frac{a^{-3}}{e^{\omega_{0}/a T_{\rm BH}}-1}.
\end{equation}
Therefore, the final relic abundance for gravitational wave can be estimated as 
\begin{eqnarray}\label{eq:omGW_PBHevap}
     h^2 \Omega_{\rm GW} &=& \frac{1}{\rho_{\rm cr,0} h^{-2}} \frac{d\rho_{\rm GW,0}}{d \ln \omega_0} \simeq   1.5 \times 10^{-33} \ \text{GeV}^{-4} \left(\frac{1 \ \rm g }{M_{\rm in}}\right)^{1/2} \omega_0^4  \ I(\omega_0),
\end{eqnarray}
where $\rho_{\rm cr, 0} \simeq  8 h^2 \times 10^{-47} \ \rm{GeV}^4$  with $h\simeq0.7$. 

We can analytically find the peak frequency at present by extremizing $h^2 \Omega_{\rm GW}$ with respect to $\omega_0$. Using the blackbody approximation for the Schwarzschild blackholes, one obtains $\omega_{\rm peak}\simeq 2.8 a_{\rm ev} T_{\rm BH}$ or more explicitly
\begin{equation}
    f_{\rm peak} \simeq 1.6 \times 10^{13} \ {\rm Hz} \left(\frac{M_{\rm in}}{1 \ \rm g}\right)^{1/2}.
\end{equation}
\begin{figure}[tbp]
    \centering
    \includegraphics[width=.5\textwidth]{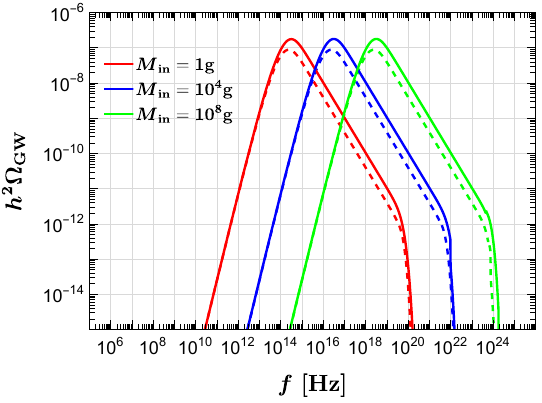}
    \caption{Gravitational wave spectrum from direct evaporation of PBH for different initial mass of PBH, $M_{\rm in} = 1 \ \rm{g}$, $10^4 \ \rm{g}$, and $10^8 \ \rm{g}$. {The solid lines in the figure represent exact numerical solution obtained using the publicly available code "ULYSSES" ~\cite{Granelli:2020pim} which incorporates the complete greybody factors into the computation without any approximation, while the dashed lines depict the semi-analytical estimate given in Eq.~(\ref{eq:omGW_PBHevap}).}}
    \label{fig:GW_direct}
\end{figure}

To evaluate the integration $I(\omega_0)$, we turn to the matter-dominated or PBH-dominated epoch, as gravitons are generated during PBH evaporation and after PBH evaporation, no further gravitons are produced. During the matter-dominated epoch, the scale factor is related to time as {$a/a_0=At^{2/3}$, where $A$ is the coefficient determined by $a/a_0 = a_{\rm ev}/a_0 \left({t}/{\tau}\right)^{2/3} = A t^{2/3}$.} Consequently using Eqs. (\ref{eq:lifetime}) and (\ref{eq:aevtoa0}), $A$ is given by:
\begin{eqnarray}\label{eq:Aquant}
     A = {\frac{a_{\rm ev}}{a_0}} \left(\frac{1}{\tau}\right)^{2/3} \simeq {1.17 \times 10^{-31}} \left(\frac{M_p^{7/6}}{M_{\rm in}^{1/2}}\right),
\end{eqnarray}

Thus the integral Eq.~(\ref{eq:Iw0}) can be written as 
\begin{eqnarray}\label{eq:Iw0withtime}
     I(\omega_0) &=& A^{-3} \int_{t_i}^{t_{\rm ev}} dt \left(1-\frac{t-t_i}{\tau}\right)^{2/3} \frac{t^{-2}}{\exp\left[\alpha t^{-2/3}\left(1-\frac{t-t_i}{\tau}\right)^{1/3}\right]-1},
\end{eqnarray}
where $\alpha\equiv\frac{\omega_0 M_{\rm in}}{A M_p^2}$. 
Typically, the frequency-dependent integral necessitates numerical evaluation for each $\omega_0$, which can then be converted to linear frequency as $f=\omega/2\pi$.
For this, we evaluate the formation time $t_i$ by using Eq.~(\ref{eq:MassPBH}), and the total energy density at the formation time of the PBH; $\rho(T_{\rm in}) = 3 M_p^2 H^2(T_{\rm in})$, as:
\begin{equation}
    t_i = \frac{M_{\rm in}}{8\pi \gamma M^2_p},
\end{equation}
and the evaporation time is approximated as $t_{\rm ev}= t_i + \tau \simeq \tau$. 

The gravitational wave spectra from direct evaporation of PBH is shown in Fig.~\ref{fig:GW_direct} as a function of frequency $f$ for different initial masses of PBH, $M_{\rm in} = 1, 10^{4}, 10^{8} \ \rm g$. {The solid lines in the figure represent exact numerical solution obtained using the publicly available code "ULYSSES"~\cite{Granelli:2020pim} which incorporates the complete greybody factors into the computation without any approximation}, while the dashed lines depict the semi-analytic results for which we assume the geometric optics limit for the evaporation of PBH. {It can be observed that the gravitational wave spectrum exhibits a change in slope after the peak. This behavior arises due to the evolution of the PBH mass over time, which is accounted for in the calculation through the integral expression given in Eq.~(\ref{eq:omGW_PBHevap}). During the Hawking evaporation process, the PBH undergoes a continuous reduction in mass and size. Consequently, as the PBH approaches the final stages of its lifetime, its temperature increases significantly. The gravitons emitted in this late phase of evaporation originate from a smaller and hotter black hole. As a result, these gravitons possess a higher initial frequency compared to those emitted earlier in the evaporation process. This leads to the formation of the high-frequency tail in the gravitational wave spectrum.
} Moreover, it is worth noticing that as the PBH mass increases, the peak frequency of the GW spectrum increases, consequently shifting the spectrum towards higher frequencies. This is due to the fact that larger mass of PBH corresponds to formation of PBH at a later time, consequently leading to a relatively shorter period of cosmological redshift thereby shifting the spectrum to higher frequencies, however, the peak amplitude remains unchanged.

\section{Gravitational Wave production from massive scalar particle}\label{section4}
Massive scalar particles of the sub-Planckian mass can be produced from PBHs at the late stage of its lifetime as a consequence of the increase in Hawking temperature as the PBH mass decreases. Once this massive scalar particle is produced from PBH, it can decay into the SM fermions or dark sector particles\footnote{{In this work, we consider only fermionic decay of $\chi$ for illustrative purpose. Bosonic decays will not change our phenomenological discussion.}}, with its lifetime being decided by its mass and couplings.
The Boltzmann equation to track the evolution of such a scalar $\chi$ can be written as   
\begin{equation}
\begin{aligned}[b]
     \frac{d \tilde{n}_\chi}{d \ln(a)} =  \frac{\tilde{\rho}_\text{BH}}{M_\text{BH}} \frac{\Gamma_{\text{BH}\rightarrow \chi}}{H} - \frac{\Gamma_{\chi \rightarrow f \bar{f}}}{H}  \tilde{n}_\chi,
     \label{Boltzeq.scalarparticle}
\end{aligned}
\end{equation}
which has to be solved along with Eq.~(\ref{Bolteq}) simultaneously.
Here $\tilde{n}_\chi \equiv n_\chi a^3$ is the comoving number density of $\chi$ and $\Gamma_\chi$ is the decay rate of scalar particle $\chi$ decaying into a pair of fermions. This decay width is given by\footnote{{The decay width for $\chi$ decaying to two scalars $\phi$ is given by : $\Gamma_{\chi\to \phi\phi}=\frac{1}{16 \pi} m_\chi (\frac{\mu}{m_\chi})^2$, where $\mu$ is the $\chi\phi^\dagger\phi$ coupling {\it i.e.} $\mathcal{L}\supset \mu \chi \phi^\dagger \phi$.}} 
\begin{equation}\label{eq:chi_decay}
    \Gamma_{\chi \rightarrow f \bar{f}} = \frac{y_f^2}{8 \pi} m_\chi,
\end{equation}
with Yukawa coupling $y_f$ and mass of scalar $\chi$ denoted as $m_\chi$.  
Thus the lifetime of $\chi$ can be estimated to be 
\begin{equation}
    \tau_\chi \simeq  7\times 10^{-26} \ {\rm s} \left(\frac{10^{-7}}{y_f}\right)^2 \left(\frac{10^{-2} M_p}{m_\chi}\right).
\end{equation}
Clearly, we can see that, by an appropriate choice of the Yukawa coupling $y_f$, the scalar $\chi$ can be long lived enough to decay after PBH evaporation {if $\tau_\chi>\tau_{\rm BH}$,
while still being safe from cosmological and phenomenological constraints. 
This requirement imposes an upper limit on the value of $y_f$ for our analysis. }

Furthermore, $\Gamma_{\text{BH}\rightarrow \chi}$ is the momentum integrated emission rate of scalar particle from PBH evaporation given  by
\begin{equation}
    \begin{aligned}[b]
        \Gamma_{\text{BH}\rightarrow \chi}(t)   =\frac{27 g_\chi}{128 \pi^3} \frac{M_p^2}{M_\text{BH}(t)}\int_{z}^\infty \frac{\psi_{\chi}(x) (x^2 - z^2)}{e^x - 1}dx,
    \end{aligned}
\end{equation}
where $x=E/T_\text{BH}$, $z = m_\chi/T_{\rm BH}$. Similar to the previous section, again resorting to the geometric optics limit for the greybody factors, we put $\psi_{\chi}(x) = 1$, which thus allows the integration to be analytically solved. Utilizing Eq.~(\ref{eq:PBHmass}) for PBH mass evolution, the emission rate is given by \cite{Cheek2022, Perez-Gonzalez:2020vnz}:
\begin{equation}\label{eq:PBHdecay}
    \Gamma_{\text{BH}\rightarrow \chi}(t) = \frac{27 g_\chi}{64 \pi^3} \frac{M_p^2}{M_\text{in}} \left(1-\frac{t-t_i}{\tau}\right)^{-1/3}\mathcal{G}(z).
\end{equation}
Here $ \mathcal{G}(z)= [z  {\rm Li}_2(e^{-z}) + {\rm Li}_3(e^{-z})]$ and $\rm{Li}_n$ is the polylog function of order $n$. It is worth noting here that for temperatures significantly higher than $m_\chi$, ({\it i.e.} $z\ll1$), $\mathcal{G}(z)$ approaches the value 
{$\zeta(3)=1.20206$} where $\zeta(n)$ is the Riemann zeta function. 
However, for our numerical calculation, we utilise the "ULYSSES" package which incorporates the modified greybody factors.

\begin{figure}[tbp]
    \centering
    \includegraphics[width=.495\textwidth]{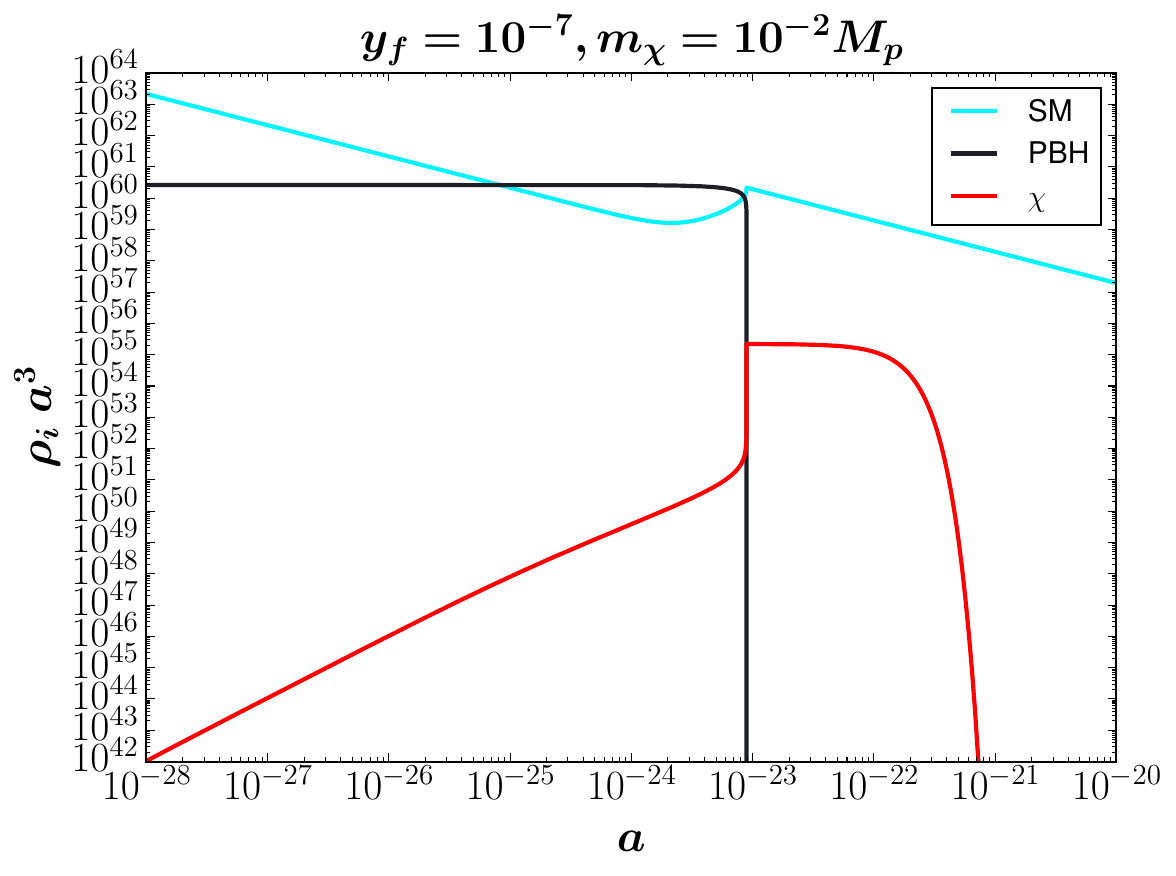}
    \hfill
    \includegraphics[width=.495\textwidth]{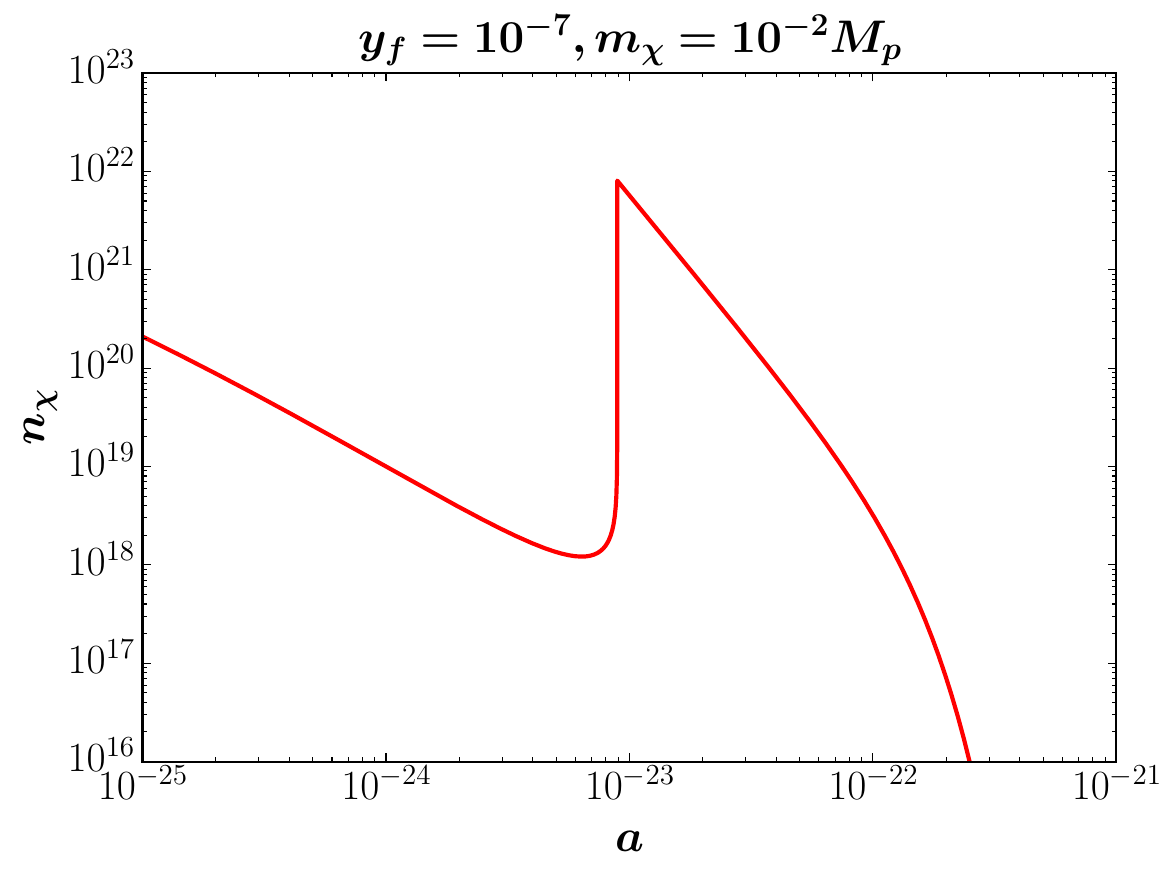}
    \caption{Evolution of the comoving energy densities for radiation, PBH and scalar particle (left) and the number density of scalar particle (right) as a function of scale factor. Here, we take the initial mass of PBH as $M_{\rm in} = 1 \ \rm{g}$.}
    \label{fig:evolution}
\end{figure}

In Fig.~\ref{fig:evolution}, the left panel shows the numerical results for the evolution of the comoving energy density of radiation (cyan), PBH (black) and the massive scalar particle (red) as a function of the scale factor $a$. One can see that PBHs dominate the energy density of the universe before they evaporate completely. The right panel shows the number density of massive scalar particle in terms of the scale factor $a$. An enhancement in the number density for $\chi$ can be noticed at the scale factor of approximately $a\sim 10^{-23}$, which corresponds to the fact that most of the massive scalar particle can be produced towards the end of PBH lifetime when $T_{\rm BH}$ increases beyond $m_\chi$ due to the PBH mass loss. 

If this produced scalar particle $\chi$ has a short lifetime such that it decays promptly into SM fermions or dark sector particles, after its production even before the end of complete evaporation of PBH, then its number density diminishes extremely quickly such that the production of the scalar $\chi$ loses its phenomenological significance as there will not be any interesting consequences. 
Conversely, we focus on the scenario where this scalar has a long lifespan and decays well after the PBH evaporation. In such a case, we can disregard the second term on the right-hand side of Eq.~(\ref{Boltzeq.scalarparticle}) for the production of scalars from PBH.
Therefore, the number density of massive scalar particle, $n_\chi$, right after PBH evaporation can be written as
\begin{equation}
    (a^3 n_\chi)|_{\rm ev} = n_{\rm BH}(t_i) a_i^3 \int_{t_i}^{t_{\rm ev}}\Gamma_{BH\rightarrow \chi}(t)\,dt,
\end{equation}
where we used the constancy of $ n_{\rm BH}(t_i) a_i^3$.
With Eq.~(\ref{eq:PBHdecay}), it becomes 
\begin{eqnarray}\label{eq:nchiatev}
    n_\chi(a_{\rm ev}) &=& \frac{27 g_\chi}{64 \pi^3}  \frac{M_p^2}{M_{\rm{in}}} n_{\rm BH}(t_i) \bfrac{a_i}{a_{\rm ev}}^3  \int_{t_i}^{t_{\rm ev}} dt \left(1-\frac{t-t_i}{\tau}\right)^{-1/3}~\mathcal{G}(z),
     \nonumber \\
    &\simeq& \frac{15~g_\chi ~\zeta(3)}{\pi^4g_*(T_{\rm BH}) } \frac{M^2_p}{m^2_\chi} n_{\rm BH}(t_i) \bfrac{a_i}{a_{\rm ev}}^3 \simeq 1.54\times 10^{21} \left(\frac{1g}{M_{\rm in}}\right)^7\left({\frac{10^{-2} M_{p}}{m_\chi}}\right)^2,
\end{eqnarray}
where we have used Eq.~(\ref{eq:PBHmass}) for PBH mass evolution and an approximation $\mathcal{G}(z)\simeq \zeta(3)$ for $t>t_1$ in the second line.
Here $t_1$ is characterized as the point when the production of $\chi$ through Hawking evaporation becomes viable {\it i.e.} when $T_{\rm BH}=m_\chi$ or $M_{\rm BH} (t_1) = {M_p^2}/{m_\chi}$ and consequently we have:
\dis{
1-\frac{t_1}{\tau} = \bfrac{M_p^2}{M_{\rm in}m_\chi}^3.
}

\begin{figure}[t]
    \centering
    \includegraphics[width=0.475\textwidth]{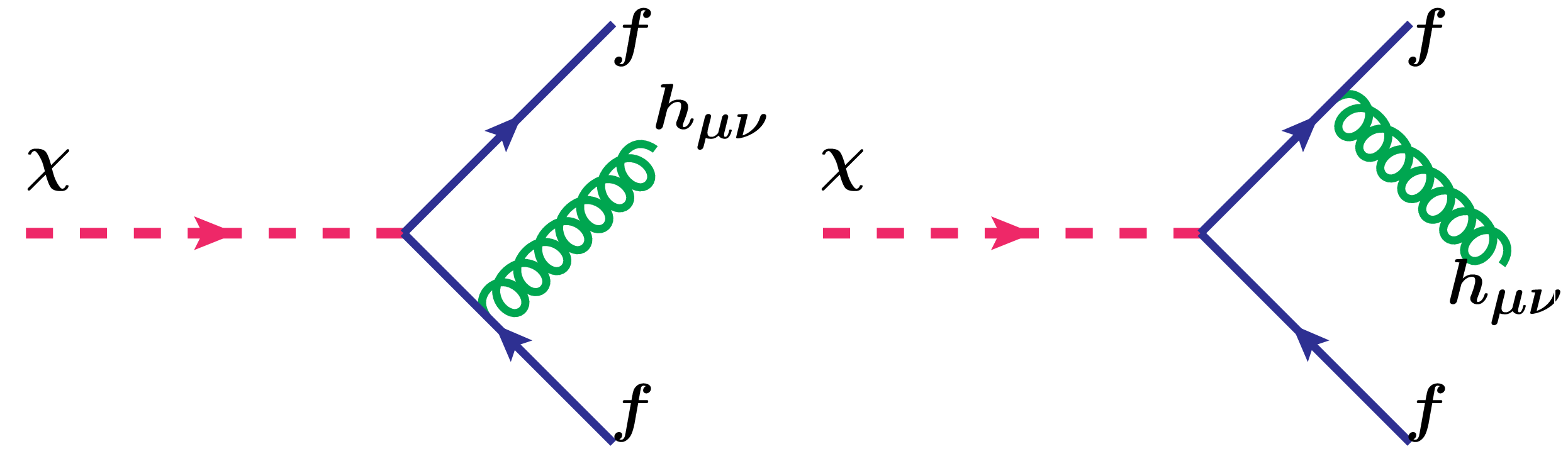}
     \includegraphics[width=0.475\textwidth]{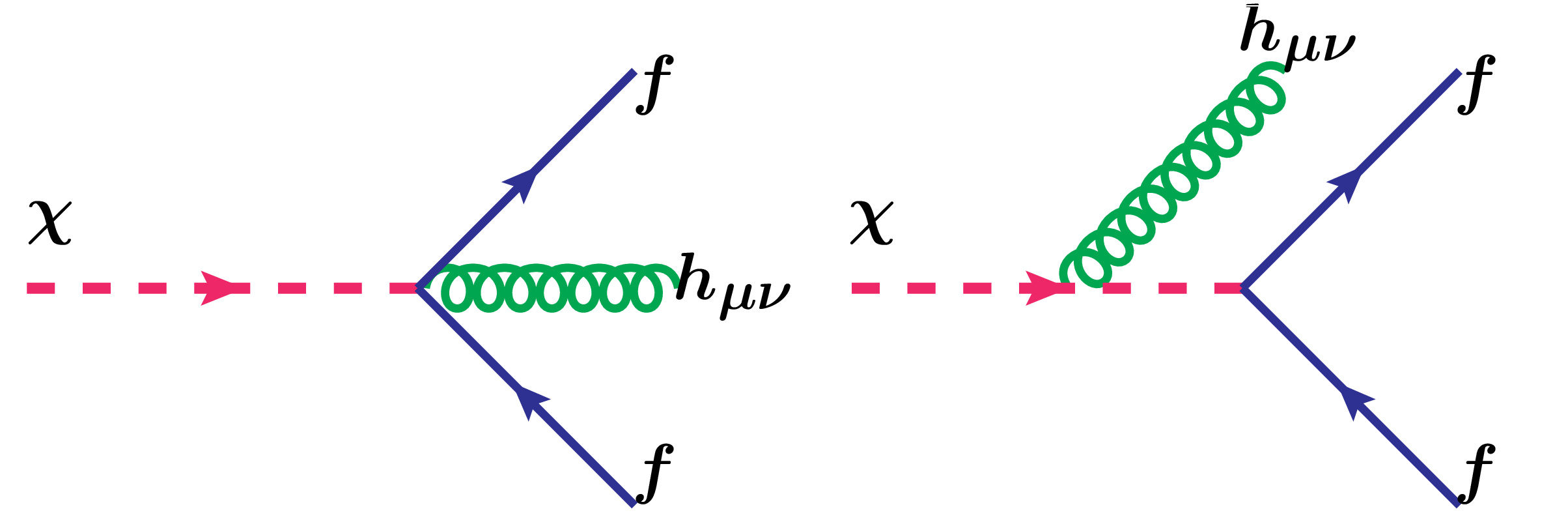}
    \caption{{Feynman diagrams representing the decay of a scalar particle ($\chi$) into fermion ($f$) along with a graviton ($h_{\mu\nu}$) bremsstrahlung.} }
    \label{fig:graviton_brems}
\end{figure}

Once these super heavy scalars are produced from the PBH evaporation, their decay into some fermions, also produces a graviton through bremsstrahlung process as depicted in Fig.~\ref{fig:graviton_brems}.
{It is important to note that the amplitudes for the last two decay processes shown in Fig.~\ref{fig:graviton_brems} are zero. Specifically, the amplitude for the third diagram, which involves graviton bremsstrahlung emission from the vertex, vanishes due to the traceless condition for a massless graviton and the amplitude for the fourth diagram, which involves graviton bremsstrahlung from $\chi$, vanishes because the decay occurs in the rest frame of $\chi$, as demonstrated in \cite{Barman:2023ymn}}. Here it is worth mentioning that, if the $\chi$ is short lived and decays immediately after its production during PBH evaporation process, then the associated energy density of the gravitons produced in the bremsstrahlung will always be subdominant to the energy density of gravitons produced directly in PBH evaporation. Thus in such a scenario, it is not possible to have a distinguishable GW signature for the scalar $\chi$. Hence, we focus on the case where $\chi$ decays much later  after PBH evaporation.

The evolution of the energy density of gravitons $\rho_{\rm GW}$  produced in the bremsstrahlung of $\chi$, can be tracked by the Boltzmann equation
\begin{equation}
    \frac{d}{dt} \left(\frac{d \rho_{\rm GW}}{dE_{\rm GW}}\right) + 4 H \frac{d \rho_{\rm GW}}{dE_{\rm GW}} = n_\chi(a_{\rm ev}) \left(\frac{a_{\rm ev}}{a}\right)^3 \frac{d \Gamma_{\chi \rightarrow \rm GW}}{dE_{\rm GW}} E_{\rm GW},
\end{equation}
or 
in terms of the scale factor $a$ as
\begin{eqnarray}
    \frac{d}{da} \left(a^4 \frac{d \rho_{\rm GW}}{d \ln E_{\rm GW}}\right) = \frac{n_\chi(a_{\rm ev}) a_{\rm ev}^3}{H} \frac{d \Gamma_{\chi \rightarrow \rm GW}}{dE_{\rm GW}} E_{\rm GW}^2,
\end{eqnarray}
where $d \Gamma_{\chi \rightarrow \rm GW}/dE_{\rm GW}$ is the differential decay rate of scalar particle. {When the decay happens after $\chi$ becomes non-relativistic, it is} given by~\cite{Kanemura:2023pnv}\footnote{{The differential decay rate of $\chi$ decaying to two scalars with the graviton bremsstrahlung can be written as: $   \frac{d \Gamma_{\chi \rightarrow \rm GW}}{dE_{\rm GW}} \simeq \frac{\mu^2 m_\chi}{128 \pi^3 M_p^2 E_{\rm GW}} (1-2 \frac{E_{\rm GW}}{m_\chi})^2$~\cite{Barman:2023ymn}}. }
\begin{equation}
    \frac{d \Gamma_{\chi \rightarrow \rm GW}}{dE_{\rm GW}} = \frac{y_f^2 m_\chi^3}{64 \pi^3 M_p^2 E_{\rm GW}} F(E_{\rm GW}/m_\chi),
\end{equation}
with form factor: $F(x) = (1-2x)(1-2x+2x^2)$. {We note that using this equation for our analysis is valid, since we focus on the late decay of $\chi$ as discussed previously. Thus, by the time $\chi$ decays and produces gravitons through the bremsstrahlung process, it has become non-relativistic.}

We consider that the generation of gravitational waves can occur from the period spanning the complete evaporation of PBH ($a_{\rm ev}$) to the decay of the scalar particle ($a_{\chi}$).

\begin{eqnarray}
     a_\chi^4 \frac{d \rho_{\rm GW}(a_\chi)}{d \ln E_{\rm GW}} &=&  \frac{y_f^2 m_\chi^3}{64 \pi^3 M_p^2} E_{\rm GW} F(E_{\rm GW}/m_\chi) n_\chi(a_{\rm ev}) a_{\rm ev}^3 \int_{a_{\rm ev}}^{a_\chi} da \frac{1}{H} \nonumber \\
     &=&  \frac{y_f^2 m_\chi^3}{64 \pi^3 M_p^2} E_{\rm GW} F(E_{\rm GW}/m_\chi) n_\chi(a_{\rm ev}) {\frac{a_{\rm ev}^3}{a_0^2}} ~\frac{2}{3 A^2} (a_\chi^3 - a_{\rm ev}^3),
\end{eqnarray}
where we use $H=\dot{a}/a$ and the relation $a/a_0 = A t^{1/2}$ with $A=a_{\rm ev}/a_0 \left(1/\tau\right)^{1/2}$ as the Universe is radiation dominated when the particle $\chi$ decays. Therefore, $A$ can be estimated to be 
\begin{equation}
    A = \frac{a_{\rm ev}}{a_0}\frac{1}{\tau^{1/2}} \simeq 7.4 \times 10^{-32} M_p^{1/2}.
\end{equation}
Now using the result of Eq.~(\ref{eq:nchiatev}), one can find the energy density of gravitational wave at $a_\chi$ as:
\begin{eqnarray}
    \frac{d \rho_{\rm GW}(a_\chi)}{d \ln E_{\rm GW}} &\simeq&  2.3 \times 10^{58}~ \frac{y_f^2 m_\chi M_p^9}{M_{\rm in}^7} E_{\rm GW} F(E_{\rm GW}/m_\chi) {\frac{a_{\rm ev}^3}{a_0^2 a_\chi}}\left[1-\left(\frac{a_{\rm ev}}{a_\chi}\right)^3\right]
\end{eqnarray}
Thus the relic abundance of gravitational wave can be estimated as 
\begin{equation}
    h^2 \Omega_{\rm GW} = \frac{1}{\rho_{\rm cr, 0} h^{-2}} \frac{d \rho_{\rm GW} (a_\chi)}{d \ln E_{\rm GW}} \left(\frac{a_\chi}{a_0}\right)^4, 
\end{equation}
where $E_{\rm GW} = 2\pi f (a_0/a_{\chi})$. Therefore, $h^2 \Omega_{\rm GW}$ can be written by

\begin{eqnarray} \label{eq:finOmGW}
     h^2 \Omega_{\rm GW} &\simeq& \frac{2.3 \times 10^{58}}{\rho_{\rm cr, 0} h^{-2}} 
     \frac{y_f^2 m_\chi M_p^9}{M_{\rm in}^7} E_{\rm GW}F(E_{\rm GW}/m_\chi){\frac{a_{\rm ev}^3}{a_0^2 a_\chi}}\left[1-\left(\frac{a_{\rm ev}}{a_\chi}\right)^3\right]
     \left(\frac{a_\chi}{a_0}\right)^4 \nonumber \\
     &\simeq& 1.8 \times 10^{105} \ {\rm{GeV}^{-4}} \frac{y_f^2 m_\chi M_p^9}{M_{\rm in}^7} F(E_{\rm GW}/m_\chi) f ~ {\bfrac{a_{\rm ev}}{a_0}^3 \left(\frac{a_\chi}{a_0}\right)^2}.
\end{eqnarray}
The temperature at which the massive scalar particle decays into fermions and thus produces the GW via bremsstrahlung, denoted as $T_\chi$, can be determined from the relation $H=1/2 \tau_\chi$, during the radiation dominated epoch as:
\begin{equation}
    T_\chi = {\left(\frac{3 \sqrt{10} y_f^2 M_p m_\chi}{16 \pi^2 g_*^{1/2}(T_\chi)}\right)^{1/2} \simeq \frac{3 y_f}{4 \pi} \left(\frac{M_p m_\chi}{g_*^{1/2}(T_\chi)}\right)^{1/2}}.
\end{equation}
From the entropy conservation, $a_{\chi}/a_0$ is given by
\begin{eqnarray}\label{eq:axtoa0}
    \frac{a_{\chi}}{a_0} &\simeq& \frac{T_0}{T_{\chi}} \left(\frac{g_{*,s}(T_0)}{g_{*,s}(T_{\chi})}\right)^{1/3} \simeq \frac{{10^{-12}}}{y_f} \left(\frac{\text{GeV}^2}{M_p m_\chi}\right)^{1/2}, 
\end{eqnarray}
where we use $ g_{*,s}(T_{\chi}) \simeq g_{*}(T_{\chi}) \simeq 100$ in the last equality.
Finally, using Eqs. (\ref{eq:aevtoa0}) and (\ref{eq:axtoa0}), we can find the relic abundance for GW as
\begin{equation}\label{eq:omGW_chi}
     h^2 \Omega_{\rm GW} \simeq 9 \times 10^{-17} \left(\frac{f}{10^{17} \ \rm Hz}\right) \left(\frac{1 \ \rm{g}}{M_{\rm in}}\right)^{5/2} F(E_{\rm GW}/m_\chi) .
\end{equation}
\begin{figure}[tbp]
    \centering
    \includegraphics[width=0.495\textwidth]{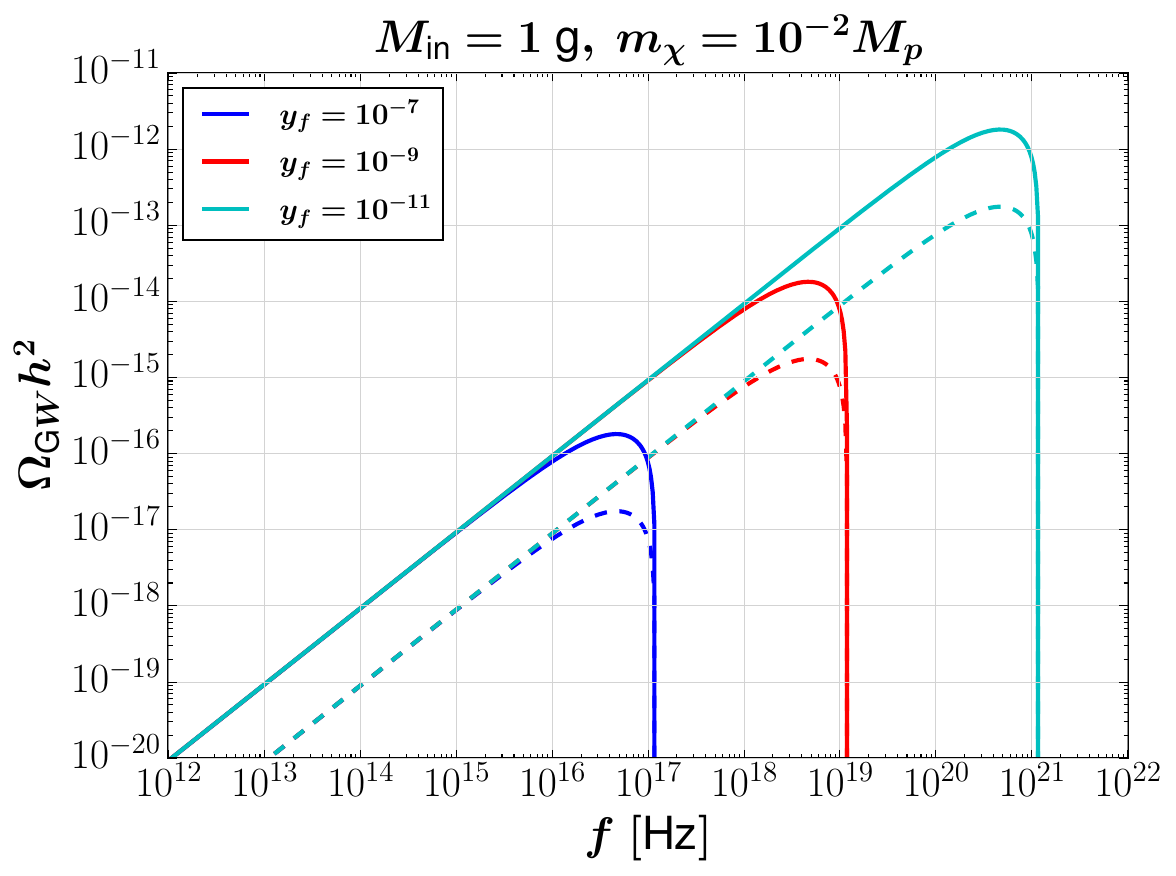}
    \hfill
    \includegraphics[width=0.495\textwidth]{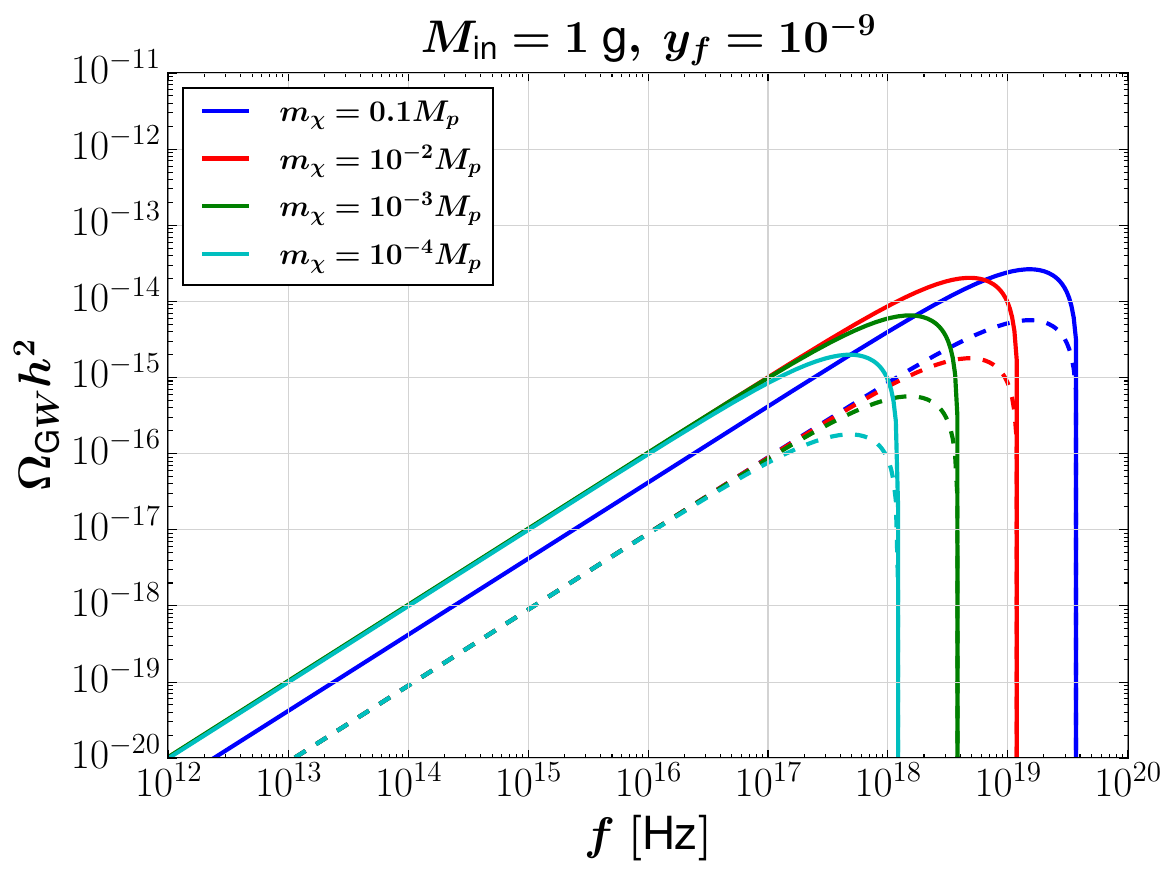}
    \caption{{Gravitational wave spectrums for graviton bremsstrahlung from the decay of massive scalar particle are shown for different values of Yukawa couplings $y_f$ when $m_\chi= 10^{-2} M_p$ (left) and for different scalar particle masses $m_\chi$ when $y_f=10^{-9}$ (right), respectively.} The numerical result (solid lines) and the analytical estimation in Eq.~(\ref{eq:omGW_chi}) (dashed lines) are shown together.}
    \label{fig:OmGWfrommassivescalar}
\end{figure}
Since we consider the GWs generated by the scalar particle decay,  the GWs at production could carry at most energy $E_{\rm GW} \leq m_{\chi}/2$. Cosequently, the peak frequency is calculated as
\begin{eqnarray}\label{eq:fpeak_chi}
    f_{\rm peak} = \frac{m_\chi}{4 \pi} \left(\frac{a_{ \chi}}{a_0}\right) 
    &\simeq& 1.2 \times 10^{17} \ \text{Hz} \left(\frac{10^{-7}}{y_f}\right) \left(\frac{m_\chi}{10^{-2} M_p}\right)^{1/2},
\end{eqnarray}
where we use the result of Eq.(\ref{eq:axtoa0}).

In Fig.~\ref{fig:OmGWfrommassivescalar}, we illustrate the amplitude of the gravitational wave (GW) spectrum, $\Omega_{\rm GW} h^2$, plotted against the frequency $f$ of the GWs {for different values of $y_f$ keeping $m_\chi$ fixed (left) and for different $m_\chi$ values for a fixed $y_f$ (right), as indicated in the plot legend. It is evident that the GW spectrum remains unaffected by different Yukawa couplings $y_f$ and scalar particle masses $m_\chi$ when $E_{GW} \lesssim m_\chi$. However, as we can see, increasing the scalar particle mass or tuning the coupling $y_f$ to smaller values, shifts the peak position towards higher frequencies. With the increase in $m_\chi$, the peak frequency also increases as higher amount of energy is being transferred to the graviton in the $\chi$ decay which can be easily understood from Eq.~(\ref{eq:omGW_chi}) and Eq.~(\ref{eq:fpeak_chi}). Additionally, tuning the coupling $y_f$ to smaller values also increases the peak frequency of the GW spectrum because smaller $y_f$ causes the scalar $\chi$ to decay at relatively late times, leading to a shorter period of cosmological redshift.}  
Moreover, we can clearly see that our numerical result and the analytical estimation are in good agreement with each other.

The dimensionless characteristic strain, $h_c$, can be defined as
\begin{equation}
    h_c = f^{-1}\sqrt{\frac{3 H_0^2}{4 \pi^2} \Omega_{\rm GW} }  \simeq 8.93 \times 10^{-19} \sqrt{\Omega_{\rm GW}h^2} \left(\frac{ \rm Hz}{f}\right),
\end{equation}
where the Hubble rate at present $H_0 = 100 h \ \text{km} \ \text{s}^{-1} \ \text{Mpc}^{-1} \simeq 3.24 \times 10^{-18} h \ \text{s}^{-1}$ .
Our result Eq.~(\ref{eq:finOmGW}) is written in terms of the dimensionless characteristic strain as
\begin{equation}
    h_c \simeq 8.5 \times 10^{-44}  \left(\frac{10^{17} \ \rm Hz}{f}\right)^{1/2}  \left(\frac{1 \ \rm{g}}{M_{\rm in}}\right)^{5/4}. 
\end{equation}

\begin{figure}[t]
    \centering
    \includegraphics[width=0.495\textwidth]{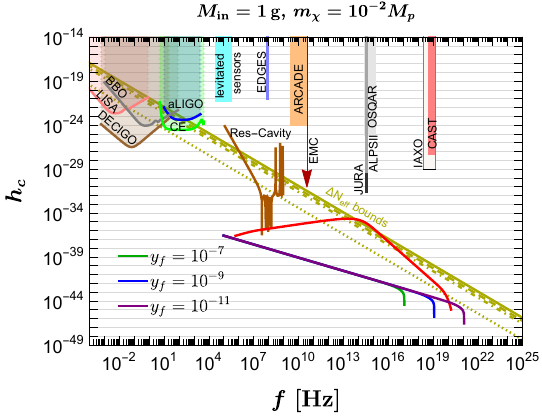}
    \hfil
    \includegraphics[width=0.495\textwidth]{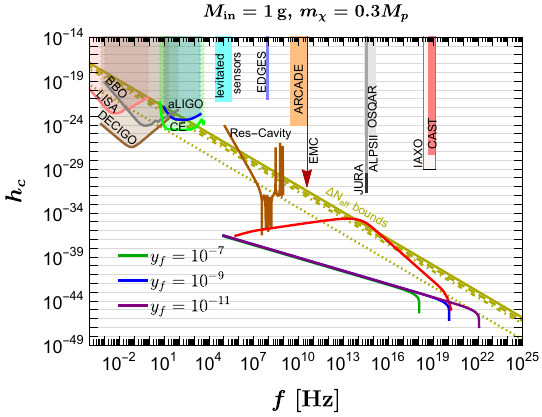}
    \caption{GW characteristic strain $h_c$ for the initial mass of PBH $M_{\rm in}=1\ \rm{g}$ and several different values of $y_f=10^{-7}, 10^{-9},$ and $10^{-11}$ with the scalar particle of mass $m_\chi=10^{-2}M_p$ (left) and $m_\chi={0.3 M_p}$ (right). Sensitivities of some current and future proposed GW  detectors~\cite{PhysRevD.88.124032, Reitze:2019iox, Seto2001, KAGRA:2013rdx, Aggarwal:2020olq} are reflected by shaded areas. The brown line illustrates the parameter space which can be probed by resonant cavities~\cite{Herman:2022fau}.  The dark yellow lines are CMB bound on $\Delta N_{\rm eff}$ from Planck~\cite{Planck:2018vyg} with solid line, and future sensitivity from CMB-S4~\cite{CMB-S4:2016ple} with dashed line, Euclid~\cite{laureijs2011euclid} with dot-dashed line,
    and proposed CMB-CVL experiment~\cite{Ben-Dayan:2019gll} with dotted line, respectively.  }\label{hc_plots}
\end{figure}

In Fig.~\ref{hc_plots}, we present the dimensionless characteristic strain $h_c$ plotted as a function of the GW frequency $f$. Here, we compare the contributions from both the direct evaporation of PBH with a red solid line and the graviton bremsstrahlung from the massive scalar particle with different values of the coupling $y_f$. Shaded regions represent the limits from various proposed GW detectors, including LISA~\cite{PhysRevD.88.124032}, the Big Bang Observer (BBO)~\cite{PhysRevD.88.124032}, CE~\cite{Reitze:2019iox}, DECIGO~\cite{Seto2001}, and the advanced LIGO~\cite{KAGRA:2013rdx}. Moreover, we also showcase the existing and aniticipated  promising experimental techniques that has the potential to detect high-frequency gravitational waves such as optically levitated sensors, enhanced magnetic conversions (EMC), inverse Gertsenshtein effect etc., including JURA, ALPS II, OSQAR, IAXO and CAST, which we adopt from~\cite{Aggarwal:2020olq}. 
The brown line illustrates the parameter space that can be probed by resonant cavities~\cite{Herman:2022fau}. It is evident that the resonant cavities with enhanced sensitivity can have the potential to test both the GW spectra. Thus, it would imply a direct probe of the Planck-scale physics, which is otherwise challenging to explore.

Energy density of GW background decreases with expansion of the universe following the behaviour of relativistic degrees of freedom $\rho_{\rm GW} \propto a^{-4}$. Thus it acts as an additional radiation field in the Universe, contributing to the effective number of neutrino species, $N_{\rm eff}$, which is defined as 
\begin{equation}
    \rho_{\rm rad} = \rho_\gamma \left(1+\frac{7}{8}\left(\frac{4}{11}\right)^{4/3} N_{\rm eff}\right),
\end{equation}
where $N_{\rm eff} = N_{\rm eff}^{\rm SM} + \Delta N_{\rm eff}$. Here $N_{\rm eff}^{\rm SM}=3.046$~\cite{Mangano:2005cc} represents the contribution from SM neutrinos and $\Delta N_{\rm eff}$ parameterizes the additional neutrino-like relativistic component. Therefore, taking $\rho_\gamma = \frac{\pi^2}{15} T_0^4$ and $\rho_{\rm GW} \simeq \Omega_{\rm GW}^{\rm max} \rho_{\rm cr,0}$, GW contribution to $\Delta N_{\rm eff}$ becomes 
\begin{equation}
    \Delta N_{\rm eff} = \frac{8}{7}\left(\frac{11}{4}\right)^{4/3} \frac{\rho_{\rm GW}}{\rho_\gamma} = \int df~ f^{-1} \Omega_{\rm GW}(f)= \frac{120}{7 \pi^2} \left(\frac{11}{4}\right)^{4/3} \frac{\rho_{\rm cr, 0}}{T_0^4} \Omega_{\rm GW}^{\rm max}.
\end{equation}

Finally, the GW relic abundance can be written in terms of $\Delta N_{\rm eff}$ as:
\begin{equation}
    h^2  \Omega_{\rm GW}^{\rm max} \lesssim 5.6269 \times 10^{-6} \Delta N_{\rm eff}.
\end{equation}

The Planck collaboration has already put a  constraint $\Delta N_{\text{eff}} < 0.30$ at 95$\%$ confidence level \cite{Planck:2018vyg}. Future experiments such as CMB-S4 \cite{CMB-S4:2016ple}, Euclid \cite{laureijs2011euclid}, and potentially cosmic-variance-limited (CVL) CMB polarization experiments are anticipated to improve the constraint on $\Delta N_{\text{eff}}$ to $\lesssim 0.06$, $\lesssim 0.013$, and $\simeq 3.1 \times 10^{-6}$, respectively \cite{Ben-Dayan:2019gll}. These constraints are depicted with dark yellow lines in Fig. \ref{hc_plots}.
The GWs originating from the direct evaporation of PBH and those stemming from the massive scalar particle bremsstrahlung decay align with the current constraints on $\Delta N_{\rm eff}$. However, they can be probed in the future CMB observations.

\section{Conclusion}\label{section5}
In this study, we have explored the production of stochastic gravitational waves not only through the direct evaporation of PBHs but also via the bremsstrahlung process during the decay of a massive scalar particle $\chi$. The GWs spectrum resulting from PBH evaporation reaches its peak frequency approximately at the Hawking temperature which tends to be considerably high for PBH with masses smaller than $\sim 10^{9}$ g as required for evaporation before the onset of BBN.  Additionally, we have examined the scenario of a long-lived, super-heavy scalar particle emerging from PBH evaporation. This particle ultimately decays into fermions, simultaneously emitting gravitons through the bremsstrahlung process. The investigation into the bremsstrahlung process as a novel source of GW production is especially intriguing for its implications in high-frequency GW scenarios. While these high-frequency GWs may be within the detection capabilities of proposed high-frequency detectors, given their sensitivity projections, they remain way below the sensitivity of the present detectors. {Nonetheless, the
distinct spectral signatures from both PBH direct evaporation and the bremsstrahlung process associated with the decay of the massive scalar particle offer promising avenues of detection by the future resonant cavity detectors.}
Moreover, GWs emanating from the decay of massive scalar particles, characterized by sub-Planckian mass scales and small Yukawa coupling, could significantly contribute to dark radiation. These contributions are expected to fall within the detection thresholds of future experiments, such as {CMB-CVL}, presenting an exciting frontier for observational cosmology.

\acknowledgments
The authors acknowledge the financial support from National Research Foundation(NRF) grant
funded by the Korea government (MEST) NRF-2022R1A2C1005050.

\bibliographystyle{JCAP}
\bibliography{reference}
\end{document}